# GroupSHAP-Guided Integration of Financial News Keywords and Technical Indicators for Stock Price Prediction

Minjoo Kim†
Graduate School of Industrial Data Engineering
Hanyang University
kmj0921@hanyang.ac.kr

Jinwoong Kim†
Graduate School of Industrial Data Engineering
Hanyang University
dnddl9456@hanyang.ac.kr

Sangjin Park*
Graduate School of Industrial Data Engineering
Hanyang University
psj3493@hanyang.ac.kr

## ABSTRACT

Recent advances in finance-specific language models such as FinBERT have enabled the quantification of public sentiment into index-based measures, yet compressing diverse linguistic signals into single metrics overlooks contextual nuances and limits interpretability. To address this limitation, explainable AI techniques, particularly SHAP (SHapley Additive Explanations), have been employed to identify influential features. However, SHAP's computational cost grows exponentially with input features, making it impractical for large-scale text-based financial data. This study introduces a GRU-based forecasting framework enhanced with GroupSHAP, which quantifies contributions of semantically related keyword groups rather than individual tokens, substantially reducing computational burden while preserving interpretability. We employed FinBERT to embed news articles from 2015 to 2024, clustered them into coherent semantic groups, and applied GroupSHAP to measure each group's contribution to stock price movements. The resulting group-level SHAP variables across multiple topics were used as input features for the prediction model. Empirical results from one-day-ahead forecasting of the S&P 500 index throughout 2024 demonstrate that our approach achieves a 32.2% reduction in MAE and a 40.5% reduction in RMSE compared with benchmark models without the GroupSHAP mechanism. This research presents the first application of GroupSHAP in news-driven financial forecasting, showing that grouped sentiment representations simultaneously enhance interpretability and predictive performance.

† These authors contributed equally
* Corresponding author

## 1. Introduction

Traditional deep learning and machine learning approaches for stock price prediction have primarily relied on structured data such as price, volume, and technical indicators, which limits their ability to incorporate unstructured factors like government policies, corporate events, and investor sentiment [1, 2]. In practice, however, unstructured elements—including policy announcements, central bank statements, geopolitical events, and social media sentiment—can substantially influence asset prices [3, 4].

For instance, on April 3, 2025, the S&P 500 index dropped by nearly 5% immediately after the U.S. president announced high tariffs, despite no changes in economic indicators or earnings reports. This highlights how market sentiment can shift sharply based on policy-driven narratives.

Recent studies have sought to quantify investor sentiment from financial news, corporate disclosures, and social media [5, 6]. A common approach converts textual information into sentiment scores (positive, neutral, negative) and combines them with technical indicators for model training [6, 7]. The development of FinBERT, a finance-specific language model, has improved sentiment classification accuracy and expanded the use of text-based features in financial forecasting.

However, this sentiment-score approach compresses diverse contextual signals into a single value, leading to information loss and reduced interpretability [7]. It also fails to reveal which specific words or topics drive model predictions.

To enhance interpretability, explainable AI (XAI) techniques have gained prominence in financial modeling [8 - 11]. In this field, interpretability is not only an academic concern but a prerequisite for reliable and transparent investment decisions [10, 11]. Among XAI methods, SHAP (SHapley Additive exPlanations) has been widely used for quantifying each feature's contribution to model output [12].

Nevertheless, SHAP suffers from exponential computational cost as it must evaluate all possible feature combinations [13, 14], making it impractical for high-dimensional text data where thousands of word-level features interact contextually.

To address this limitation, this study proposes a GroupSHAP-based approach that groups input variables at the word, sentence, or document level to compute contribution scores. Financial news and social media texts contain vast vocabularies with dynamic contextual dependencies, making it difficult to directly interpret individual token-level impacts.

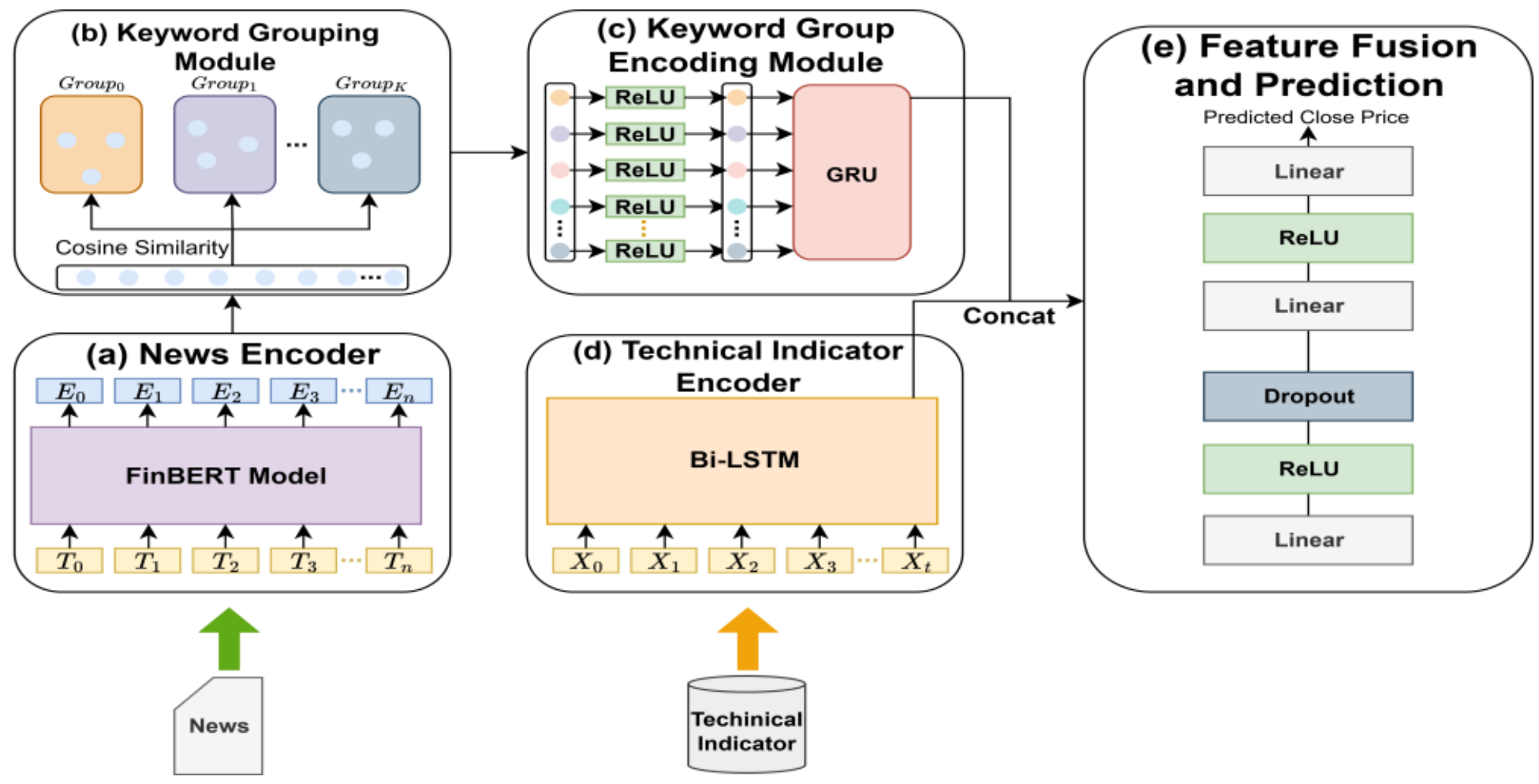

**Figure 1: Architecture of the GroupSHAP.**

However, our GroupSHAP mechanism resolves this by quantifying group-level Shapley values, thereby improving both efficiency and interpretability. Specifically, FinBERT-based sentiment embeddings are clustered into semantically coherent groups using cosine similarity, allowing the model to capture topic-driven sentiment flows linked to market reactions. These grouped sentiment features are integrated with technical indicators in a GRU-based time-series model, where GroupSHAP is applied to measure the contribution of each group.

Unlike traditional SHAP, which struggles with token-level complexity, GroupSHAP aggregates features into interpretable semantic units, achieving a balance between computational scalability and explanatory depth, and extending model interpretability to real-world market event analysis [20, 21].

The expected outcomes of this study are as follows:

- We demonstrate that collective sentiment expression not only improves forecast accuracy but also provides interpretable insights into how collective market sentiment influences price dynamics.
- By integrating explainable AI with financial text modeling, this study bridges the gap between predictive performance and interpretability, providing a practical framework applicable to risk management and market monitoring.

The remainder of this paper is organized as follows. Section 2 reviews the related literature. Section 3 presents the proposed methodology, including data preprocessing, semantic grouping, and model architecture. Section 4 reports the experimental results and performance analysis. Section 5 concludes the paper with key findings and implications.

## 2. Related Work

Early opinion-based studies focused on quantifying the relationship between news and investor sentiment. Some demonstrated that the tone of news articles could explain fluctuations in stock returns and trading volumes, while others showed that collective moods derived from social media could contribute to predicting market indices [3, 15]. These approaches empirically demonstrate the correlation between public opinion and financial markets, laying the groundwork for subsequent sentiment analysis-based forecasting research.

Subsequent studies integrated social media sentiment into stock prediction using machine learning and deep learning techniques. For example, an approach combining topic modelling and sentiment analysis demonstrated that sentiment extracted by topic from online posts is effective for predicting stock price directionality. Furthermore, various deep learning architectures such as CNN, LSTM and GRU models have been applied to financial forecasting, achieving higher predictive performance and profitability compared to traditional techniques [16, 17].

More recently, studies have employed finance-specific language models to incorporate textual sentiment into price prediction. Some used FinBERT to derive sentiment probabilities (positive, negative, neutral) for each news article, averaged them into daily sentiment vectors, and combined these with LSTM-based models to reduce prediction errors.

Other research enhanced S&P 500 forecasting by integrating ESG-related sentiment indicators and technical indicators into Bi-LSTM model, demonstrating that sentiment-rraware models outperform purely technical ones [18, 19].

In parallel, studies employing GRU-based architecture have also demonstrated superior performance in capturing delayed emotional effects, showing improved convergence speed and predictive stability compared to LSTM model in financial time-series tasks.

These findings suggest that combining finance-specific language models (e.g., FinBERT) with efficient sequential models such as GRU model can better capture temporally lagged sentiment impacts in market dynamics.

Furthermore, rather than using sentiment signals and technical or macroeconomic indicators directly as independent features, it is necessary to interpret and organize them into meaningful groups for improved explainability. In this context, GroupSHAP defines the Shapley value of "meaningful feature groups," simplifying both computation and interpretation in high-dimensional, highly correlated settings.

## 3. Methodology

This study proposes a framework for predicting the S&P 500 index by integrating news-based sentiment data with technical indicators and enhancing model interpretability through the GroupSHAP technique.

Figure 1 illustrates the overall architecture of the proposed framework, which consists of four main modules: (i) data collection and preprocessing, (ii) keyword grouping, (iii) group encoding with GRU-based modeling, and (iv) feature fusion and prediction. Each of the following subsections (3.1–3.4) corresponds to a specific component of this architecture, explaining its role and implementation in detail.

### 3.1 Data Collection and Preprocessing

This stage provides the raw inputs for both the News Encoder (Figure 1.a) and Technical Indicator Encoder (Figure 1 d) shown in Figure 1.

News articles (2015–2024) and financial market data were collected via web crawling and Yahoo Finance, respectively.

Daily financial news articles related to the S&P 500 were retrieved from Google News using targeted queries. Full texts were extracted through HTML parsing, then deduplicated and filtered to retain only those containing finance-related terms such as "*market*", "*economy*", and "*stock*". Approximately 2,500 distinct articles remained after filtering, all of which were aligned to U.S. trading days to ensure temporal consistency with market data.

Following prior financial forecasting studies that integrate both technical and macroeconomic indicators [1, 4, 15], this study employs a comprehensive set of features to capture diverse drivers of stock price movements.

Specifically, twenty variables were selected, as follows: (i) commonly adopted technical indicators—such as trading volume, moving averages (SMA, EMA), RSI, MACD, and Bollinger Bands—that capture short- and medium-term momentum and volatility patterns [1, 16], and (ii) macroeconomic and alternative asset indicators—including gold, Bitcoin, and WTI prices, bond yields (2-year and 10-year U.S. Treasury), and the U.S. dollar index—that reflect investors' risk appetite, hedging demand, and cross-market capital flows [4, 6].

The closing price served as the prediction target (Y) and was aligned on a daily basis with all input variables.

To capture short-term market reactions to external information, the prediction target was defined as the next-day closing price, meaning that all models were trained to forecast one trading day ahead.

All input features were normalized using a MinMaxScaler fitted on the training data to ensure consistent scaling between variables.

### 3.2 Keyword Grouping Module

This subsection corresponds to the Keyword Grouping Module (Figure 1 b) in the overall architecture.

For each news document, sentiment probabilities (positive, negative, neutral) were extracted using FinBERT-based News Encoder (Figure 1 a) and aggregated into daily average sentiment scores. Subsequently, word and sentence embeddings were clustered into five semantically similar groups using cosine similarity-based clustering.

The resulting features—average sentiment embeddings of each group (group_0_weight to group_4_weight)—were used as input variables.

### 3.3 Keyword Group Encoding Module

This subsection corresponds to the Keyword Group Encoding Module (Figure 1 c). This study employed the GRU as the sole model for stock prediction. GRU possesses structural advantages that enable it to efficiently learn long-term dependencies in time-series data, while maintaining sufficient expressiveness through a simplified gate structure that ensures computational efficiency compared to LSTM model.

Given that emotional factors in financial markets—such as policy announcements, news, and investor sentiment—tend to be reflected with a temporal lag rather than immediately [23], GRUs is considered well-suited for modeling such delayed effects [22].

Two GRU-based models with different input feature sets were compared:

- **GRU with technical indicators**: uses only technical indicators as input
- **GRU with full indicators**: incorporates technical indicators, FinBERT-based sentiment scores, and semantic group features

This comparison above aims to evaluate the contribution of sentiment information and grouped features to predictive performance.

### 3.4 Feature Fusion and Prediction

This stage corresponds to the Feature Fusion and Prediction Module (Figure 1 e), where the outputs from the Keyword Group Encoder (Figure 1 c) and the Technical Indicator Encoder (Figure 1 d) are concatenated and passed through a stack of linear and ReLU layers for final price prediction.

Model performance was assessed using MAE, RMSE, $R^2$ and MAPE. Additionally, the GroupSHAP method was employed to quantitatively analyze the contribution of each sentiment group feature to the model's predictions. GroupSHAP defines the contribution of a particular group $g \in G=\{1,\dots,n_G\}$ as follows:

$$\phi_g = \sum_{S\subseteq G\setminus\{g\}} \frac{|S|!(n_G-|S|-1)!}{n_G!}\left[v(S\cup\{g\}) - v(S)\right] . \qquad (1)$$

Here, $\boldsymbol{v}(S)$ denotes the expected prediction value (or reduction in prediction error) when only group S is activated. By reducing the combinatorial space from $2^M$-1 for individual features to $2^{n_G}-1$ for grouped features, GroupSHAP enhances computational efficiency and enables interpretable analysis at the group level, where $M$ is the total number of individual features and $n_G$ is the number of feature groups.

# 4. Experiments and Results

## 4.1 Sensitivity Analysis of Group Number

To address the question of how the number of semantic groups affect predictive performance, a sensitivity analysis was conducted by varying $n_G$ from 1 to 9. Cosine similarity–based clustering was applied to FinBERT embeddings to form semantically coherent groups. As summarized in Table 1, model performance (MAE, RMSE, $R^2$) was evaluated for each configuration using the GRU with full indicators model.

The results indicate that five groups ($n_G$ = 5) provide the highest explanatory power ($R^2$ = 0.7300) while maintaining stable MAE and RMSE values. Configurations with fewer than three groups oversimplified the semantic structure, losing meaningful contextual distinctions, whereas configurations with more than six groups resulted in performance degradation and instability due to over-segmentation.

Therefore, the five-group configuration was adopted for all subsequent experiments as the optimal trade-off between semantic granularity, model simplicity, and computational efficiency.

**Table 1**. Performance comparison across group configurations

| Number of Groups ($n_G$) | MAE | RMSE | $R^2$ |
| --- | --- | --- | --- |
| 1 | 82.4309 | 98.8115 | 0.6659 |
| 2 | 119.7635 | 137.7331 | 0.4446 |
| 3 | 111.2796 | 129.0076 | 0.5719 |
| 4 | 91.1171 | 106.0464 | 0.5871 |
| **5** | **85.1086** | **101.7042** | **0.7300** |
| 6 | 144.9562 | 165.1993 | 0.3875 |
| 7 | 105.3042 | 122.4136 | 0.5156 |
| 8 | 110.0346 | 127.0528 | 0.4668 |
| 9 | 103.0801 | 120.9002 | 0.6786 |

*Note:* The five-group configuration (highlighted in bold) achieved the highest explanatory power ($R^2$ = 0.7300) and balanced error levels (MAE = 85.1, RMSE = 101.7), confirming it as the optimal setting for subsequent experiments.

## 4.2 Comparison between GroupSHAP and Token-level SHAP

To evaluate the computational implications of incorporating the GroupSHAP interpretability module, an empirical comparison was conducted between token-level and group-level SHAP computations under identical experimental conditions.

While the additional interpretability stage introduces a post-hoc computation cost, the proposed grouping strategy significantly improves scalability for real-world financial prediction tasks.

As summarized in Table 2, the token-level SHAP analysis—applied to the full feature set comprising approximately 320 input variables (including FinBERT-derived keywords and technical indicators) required 9,804.7 seconds (≈163 minutes) per rolling window.

In contrast, the group-level SHAP computation with five aggregated semantic groups completed in 1,524.6 seconds (≈25 minutes), representing an 84.5% reduction in wall-clock time.

This result demonstrates that semantic grouping mitigates the combinatorial explosion of feature attribution from $2^M - 1$ to $2^{n_G} - 1$ subset without compromising interpretability, providing a scalable and computationally efficient explanation mechanism for high-dimensional financial models that integrate both textual and technical features.

**Table 2.** Comparison of SHAP Computation Costs between Token-Level and Group-Level Settings

| Method | Features | Computation Time (sec) | Time (min) | Reduction (%) |
| --- | --- | --- | --- | --- |
| Token-level SHAP (baseline) | 320 | 9,804.7 | 163.4 | – |
| GroupSHAP (proposed) | 28 | 1,524.6 | 25.4 | **84.5 ↓** |

*Note*: Experiments were executed on a single NVIDIA A100 GPU (PyTorch 2.2.0) under identical model and batch settings.

## 4.3 Overall performance comparison

The prediction models are evaluated using a rolling-window scheme. Each model was trained over a 3-year window and evaluated on the subsequent 1-year period (e.g., trained on 2015–2017 and evaluated on 2018, then trained on 2016–2018 and evaluated on 2019), ensuring a fair year-by-year performance comparison. In addition, the forecasting horizon was set to one year ahead.

Based on the test results from 2018 to 2024, the GRU with full indicators model consistently outperformed the GRU with technical indicators model in most years, indicating the added predictive value of incorporating sentiment features and semantic group embeddings.

**Table 3**. Model Performance Comparison from 2018 to 2024.

| Year | HV (%) | Tech only | | | | Full (Tech and GroupSHAP) | | | |
| --- | --- | --- | --- | --- | --- | --- | --- | --- | --- |
| | | MAE | RMSE | MAPE | $R^2$ | MAE | RMSE | MAPE | $R^2$ |
| 2018 | 32.4 | 36.4 | 45.1 | 1.33 | 0.81 | 34.3 | 44.6 | 1.27 | 0.81 |
| 2019 | 24.2 | 118.3 | 133.2 | 3.97 | 0.17 | 106.8 | 119.8 | 3.58 | 0.22 |
| 2020 | 51.7 | 78.3 | 100.9 | 2.55 | 0.90 | 99.6 | 119.9 | 3.20 | 0.86 |
| 2021 | 40.8 | 87.1 | 107.5 | 1.97 | 0.85 | 65.4 | 80.2 | 1.51 | 0.92 |
| 2022 | 73.8 | 184.3 | 211.2 | 4.55 | 0.34 | 163.1 | 190.5 | 4.01 | 0.50 |
| 2023 | 45.5 | 46.2 | 56.1 | 1.08 | 0.94 | 73.4 | 88.1 | 1.07 | 0.84 |
| 2024 | 48.8 | 99.8 | 114.4 | 1.80 | 0.89 | 53.3 | 68.9 | 0.96 | 0.96 |

*Note*: Since 2020, the performance gap between the two models has widened significantly, with GRU with full indicators achieving substantial performance improvements particularly in 2024. HV is the historical annualized volatility, calculated as $HV = \sigma_{\text{daily}} \times \sqrt{252}$, where $\sigma_{\text{daily}}$ is the standard deviation of daily log returns.

## 4.4 Analysis of Recent Results

The year 2024 represents the most recent and volatile period in the financial markets during the study. According to the annualized historical volatility (HV) statistics summarized in Table 3, the S&P 500 exhibited an HV of 48.81% in 2024, which was nearly twice as high as the relatively stable pre-pandemic period of 2018–2019, which had an average HV of approximately 28.3%.

The sharp increase in volatility highlights a structural shift in market dynamics during the post-pandemic adjustment period, characterized by heightened uncertainty and rapid price fluctuations. Consequently, 2024 provides a rigorous stress-testing environment for evaluating the robustness of the sentiment-integrated model under unstable market conditions. The forecast results for this period are summarized as follows:

- **GRU with technical indicators**: MAE = 99.78, RMSE = 114.42, $R^2$ = 0.89
- **GRU with full indicators**: MAE = 53.33, RMSE = 68.90, $R^2$ = 0.96

In other words, compared to the model using only technical indicators, GRU with full indicators achieved a 32.2% reduction in MAE, a 40.5% reduction in RMSE, and a 7% improvement in $R^2$. These results demonstrate a statistically significant improvement in predictive performance.

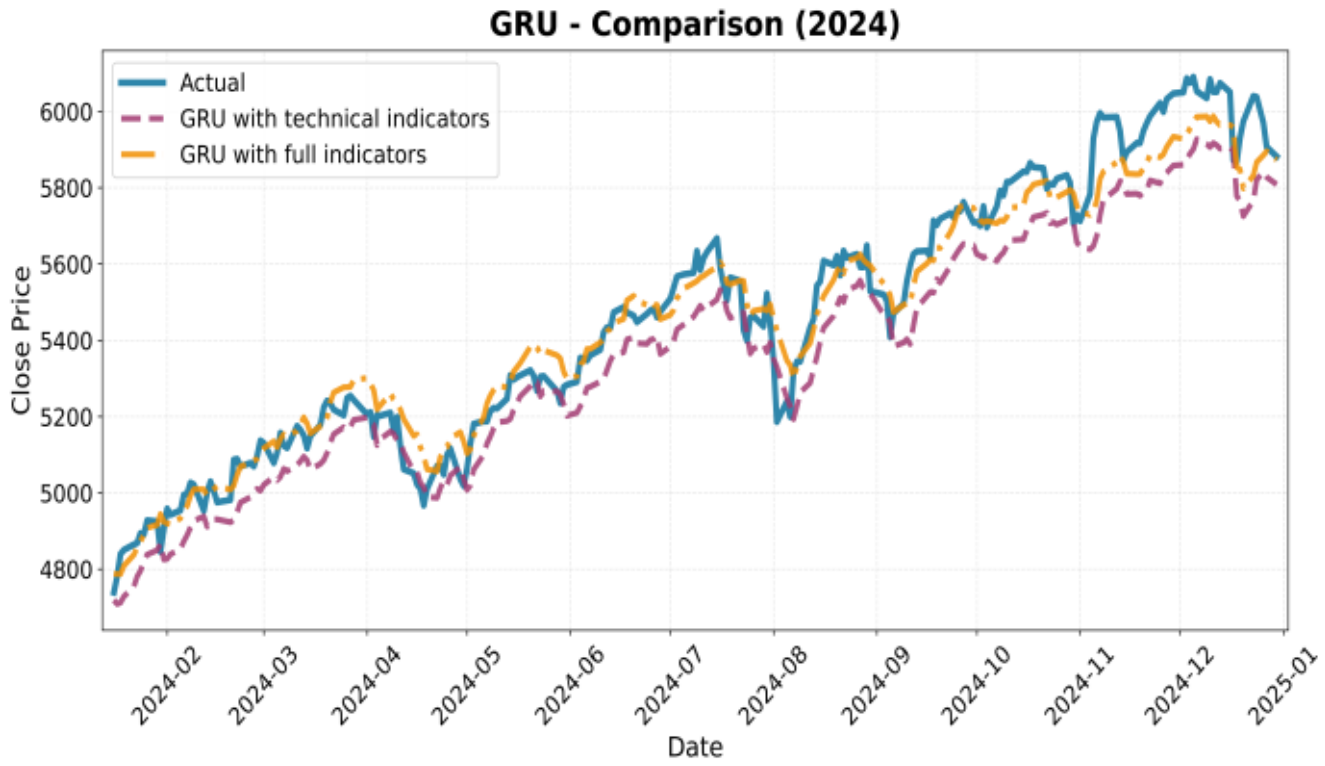

**Figure 2.** Comparison of Actual and Predicted Stock Prices in 2024

Figure 2 above compares actual stock prices in 2024 with the predictions from the two models.

- **GRU with technical indicators,** was able to track the long-term trend but exhibited significant errors during periods of high volatility (i.e., sharp rises and falls).
- **GRU with full indicators**, on the other hand, more accurately captured both the trend and fluctuations of the actual stock prices, with notably reduced errors just before sharp declines or rebounds.

This suggests that the news-based sentiment group features served as important signals for anticipating market turning points.

### 4.5 Experimental Settings

All experiments were implemented in Python (PyTorch 2.2.0) and executed on a single NVIDIA A100 GPU.

The GRU model employed two hidden layers with 256 units, a dropout rate of 0.1, and ReLU activation between fully connected layers. The AdamW optimizer was used for model training with a learning rate of $1\times10^{-4}$. The batch size was 32, time step length was 10, and training was performed for 50 epochs without separate validation, consistent with prior time-series forecasting studies.

## 5. Conclusion and Future work

This study proposes a novel framework that integrates a GRU-based stock prediction model with unstructured news data, utilizing semantic group-level sentiment features and the GroupSHAP interpretability method. Unlike previous studies that relied on simple average sentiment scores—leading to information loss—our approach enhances both the richness of sentiment representation and the interpretability of the model.

Experimental results, particularly in the 2024 prediction task, show that the GRU with full indicators model achieved substantial improvements over GRU with technical indicators, with a 32.2% reduction in MAE, a 40.5% reduction in RMSE, and an $R^2$ of 0.96. These findings demonstrate that sentiment group features are not merely auxiliary variables but make meaningful contributions to predicting market volatility and directional movements.

To the best of our knowledge, this is the first study to apply GroupSHAP to public opinion-based financial prediction. It demonstrates the feasibility of achieving both predictive accuracy and interpretability in unstructured text environments. Furthermore, the use of the GRU model—capable of capturing temporal accumulation effects and delayed responses—was reaffirmed as particularly well-suited for modeling sentiment flows in financial markets.

Nonetheless, this study is still in an early stage. It has limitations in terms of parameter optimization, such as the number of sentiment tokens, number of groups, and clustering methods. Moreover, further benchmarking against a wider range of baseline models (e.g., LSTM, Temporal Fusion Transformer, Transformer-based architectures) remains a task for future research.

In future work, we aim to extend the current framework toward a more comprehensive "semantic reconstruction and quantitative interpretation of sentiment information." In particular, our goal is to empirically verify how grouped sentiment representations provide interpretable insights into how collective market sentiment influences price dynamics. Moreover, by integrating explainable AI with financial text modeling, future research will further bridge the gap between predictive performance and interpretability, ultimately contributing to practical applications in risk management and market monitoring.

## ACKNOWLEDGMENTS

This work was supported by the National Research Foundation of Korea(NRF) grant funded by the Korea government(MSIT)(No. RS-2025-00554384), and the Technology Development Program (No. RS-2024-00513926) funded by the Ministry of SMEs and Startups (MSS, Korea).